\documentclass[aps,twocolumn,groupedaddress,superscriptaddress,showpacs,prb,floatfix]{revtex4}

\usepackage{graphicx}
\usepackage{dcolumn}
\usepackage{bm}
\usepackage{epsfig}
\usepackage{dsfont,psfrag}
\usepackage{color}

\newcommand{\be}{\begin{eqnarray}}
\newcommand{\ee}{\end{eqnarray}}
\def\refeq#1{(\ref{#1})}
\renewcommand{\vec}[1]{\boldsymbol #1}

\def\wt{\widetilde}

\def\L{\Lambda}

\def\Or{\mathcal O}

\def\al{\alpha}

\def\l{\left}
\def\r{\right}

\def\te{\mbox{e}}
\def\rmi{{\rm i}}

\def\up{\uparrow}

\def\down{\downarrow}
\def\Down{\Downarrow}

\begin{document}
\bibliographystyle{amsplain}
  \title{Dynamics and decoherence in the central spin model using exact methods}
\author{Michael Bortz}
\email{michael.bortz@itwm.fraunhofer.de}
\affiliation{Physics Department and Research Center OPTIMAS, Technische Universit\"at Kaiserslautern, 67663 Kaiserslautern, Germany}
\affiliation{Fraunhofer ITWM, 67663 Kaiserslautern, Germany}
\author{Sebastian Eggert}
\affiliation{Physics Department and Research Center OPTIMAS, Technische Universit\"at Kaiserslautern, 67663 Kaiserslautern, Germany}
\author{Christian Schneider}
\affiliation{IfB, HIF E12, ETH H\"onggerberg, 8093 Z\"urich, Switzerland}
\author{Robert St\"ubner}
\affiliation{Technische Universit\"at Dortmund, Fakult\"at Physik,  44221 Dortmund, Germany}
\author{Joachim Stolze}
\affiliation{Technische Universit\"at Dortmund, Fakult\"at Physik,  44221 Dortmund, Germany}

\date{\today}
\begin{abstract}
The dynamics and decoherence of an electronic spin-1/2 qubit  coupled to a bath of 
nuclear spins via hyperfine interactions in a quantum dot is studied.
We show how exact results from the 
integrable solution can be used to understand the dynamic behavior
of the qubit.
It is possible to predict the main frequency contributions and their broadening  
for relatively general initial states analytically,
leading to an estimate of the corresponding decay times, 
{which are related to $T_1$ of the electron.}
Furthermore, for a small bath polarization, a new {low-frequency} 
time scale is observed.  
\end{abstract}
\maketitle
One of the most {challenging tasks in both} theoretical and
experimental {studies of quantum information processing} is the
understanding and control of {decoherence} due to the interaction with
the environment. A prototypical example {for the loss of quantum
  information} is an electron trapped in a quantum dot, surrounded by
nuclear spins.  On short time scales up to 1ms, \cite{joh05} the
Heisenberg exchange resulting from the hyperfine interaction between
the electron and the nuclei {dominates}, before spin-orbit coupling or
dipole-dipole-interactions between the bath spins become effective.
\cite{kha00,sch03}  {
  Indeed, {impressive} experimental progress has been made over the
  recent years to observe {and control} oscillations {of single
    electron spins} coupled to bath spins {by various techniques.}
  \cite{jel04,ber08,pre08,han08,grei09}  
{In a typical magnetic resonance experiment,
    for example, one applies a static magnetic field creating a Zeeman
    splitting of the spin's energy levels, plus a resonant or near
    resonant alternating field transverse to the Zeeman field. 
    Spin rotation angles can then be controlled
    e.g. by the duration of the transverse field pulse. The
    decoherence processes on which we focus here, however, are}
not directly linked to the applied fields,
but are instead caused mainly by the Heisenberg coupling to the nuclear bath spins, which is
given by 
Hamiltonian 
\be
H=\sum_{j=1}^{N_b}A_j \vec S_0 \cdot \vec S_j 
\label{gaudef}\,,
\ee
where $N_b$ is the number of bath spins and $\vec S_0$ is the central electron spin.
This so called central spin model
is one of the most studied theoretical models for decoherence,
\cite{rev,kha02,sch02,dob03,coi04,erl04,has06,coi06,den06,che07} which 
can also be treated by exact methods.\cite{gaubook,bor_inhom07,bor_static_09}
In this paper we are interested in an analytic understanding of the 
detailed {decoherence-induced dynamics} of $\langle S_0^z\rangle(t)$
due to the coupling to the spin bath.
We therefore do not consider any external magnetic fields, but instead specify the
initial overall polarization of the system, which is conserved.
The central spin $\vec S_0$ is initially assumed to be in its down state, 
independent of the bath spins. This initial product state gets 
entangled by the exchange interactions, leading to the decoherence of the central spin.}

{The  couplings $A_j>0$ in Eq.~(\ref{gaudef}) 
are proportional to the square of the electronic wave
function at the position of the nucleus $j$. 
The methods we apply in the following do not depend on the special choice of the $A_j$. 
For definiteness, we assume a 
Gaussian distribution with the site index (distance) $j$
\cite{coi04,bor_static_09} 
\be
A_j=\frac{x_1 N_b\exp\l[-( jB/N_b)^2\r]}{\sum_{j=1}^{N_b}\exp\l[-( jB/N_b)^2\r]}\label{coup},
\ee
which allows an easy control over the two relevant characteristics 
of the distribution of $A_j$, 
namely the mean 
value $x_1=\frac{1}{N_b}\sum_j A_j$ and the degree of inhomogeneity as parametrized by $B$.
Generally the results are largely insensitive to the overall shape of the distribution, (e.g. 
in case a higher dimensional site index is used) as long as the mean $x_1$ and the
degree of inhomogeneity  are the same.}

{For homogeneous couplings, $A_j\equiv A\,\forall j$,
a non-trivial time scale $\tau {\sim A^{-1} N_b^{-1/2}}$ has been identified using exact 
methods,\cite{mbjshom07,koz07} which can be
{interpreted} as a decoherence time.
A number of 
authors \cite{kha02,sch02,sch03,dob03,coi04,erl04,has06,coi06,den06,che07} have
studied the influence of inhomogeneous couplings by a variety of approximate 
numerical and analytical methods.  
Obviously the use of exact Bethe Ansatz methods to
the general dynamic problem would be a great advantage, but so far
the possibilities have been limited
to certain
non-equilibrium situations in
the related BCS-model \cite{caux09}
and to the special case of a maximal bath polarization.
\cite{bor_inhom07} }
Here we demonstrate how Bethe Ansatz results can be used to obtain the central 
features of the 
dynamic behavior for more general polarizations and coupling constants 
and compare with numerical complete diagonalization results.

We first want to consider 
{an initial state $|L\rangle=|\Down,\up,\up,\down,\ldots,\up\rangle$, where the
central spin $\Down$ and $M_b$ bath spins at 
specified sites $L=\left\{\ell_1,\ldots,\ell_{M_b}\right\}$ are in the down state.
All other spins are in the up state.  Hence $|L\rangle$ is an eigenstate
of all $S_i^z$ operators with total magnetization $S^z_{\rm tot}=N_b/2-M_b-1/2$ and
is initially not entangled in any way. }
The time evolution is given in terms of the 
eigenstates $|M_\nu\rangle$ and eigenvalues $\L_{M_\nu}$
of the model \refeq{gaudef} 
{
\be
|L(t)\rangle&=&\sum_{\nu}\te^{-\rmi \L_{M_\nu} t}|M_\nu\rangle\langle M_\nu|L\rangle\label{lt}\; .
\ee  }
{From \refeq{lt} an explicit expression for the reduced density matrix of the central spin can be derived,\cite{bor_inhom07} which we employ to evaluate}
\be
\langle S_0^z\rangle(t)=\frac12\l(1-2 \sum_{J}^{C_{M_b}^{N_b}}|\al_{J}(t)|^2\r)\label{sz} 
\ee
with
{
\be
\al_{J}(t)&=&\langle J|L(t)\rangle = \sum_{\nu}
 \langle J|M_\nu\rangle\langle 
M_\nu|L\rangle
\te^{-\rmi t \L_{M_\nu}}
\label{aljb},
\ee}
where  both $|L\rangle$ and 
$|J\rangle$ are {eigenstates of all $S_i^z$} with 
the central spin fixed in the down state. Thus it suffices to specify 
only the $M_b$ flipped bath spins in the subsector of
dimension $C_{M_b}^{N_b}=N_b!/((N_b-M_b)!M_b!)$.
In Ref.~[\onlinecite{bor_inhom07}] the matrix {elements $\langle M_\nu|L\rangle$ 
were} {explicitly} given in terms of {the quantum numbers (Bethe roots, see below) of the energy eigenstates $| M_\nu \rangle$ and of the {$S_i^z$ eigenstates} $|L\rangle$}; 
alternatively, they can be obtained from a complete diagonalization. 
However, an exact calculation of the large number of terms in the sums (\ref{sz}) and 
(\ref{aljb}) is impossible already for modest system sizes, except for special cases.
In particular, a fully polarized bath $S^z_{\rm tot}=N_b/2-1/2$ (i.e.~$M_b=0$)
was studied in Refs.~[\onlinecite{kha02,bor_inhom07}],
where the sum (\ref{aljb}) only contains $N_b+1$ terms. 

We now would like to consider a more general polarization{, $M_b \neq 0$,} and 
single out the most important contributions to the sum (\ref{aljb}).
In this way, it is possible to estimate the dominant frequency scales and the
widths of peaks in 
{the frequency spectrum of Eq.~(\ref{sz})} and thus to obtain the decoherence time.
Our strategy is based on results in Ref.~[\onlinecite{bor_static_09}], where it
was found that only a few product states $|J\rangle$ have an appreciable 
overlap $\langle J|M_\nu\rangle$ for a given $|M_\nu\rangle$ as long as $M_b\ll N$.  
These product states are essentially those obtained from the classical
ground state 
{$| \Downarrow, \uparrow, \uparrow,...,\uparrow\rangle $} {by flipping certain 
individual nuclear 
bath spins} as outlined below, {that spin pattern being} also reflected in the local
{expectation values $\langle S_j^z\rangle$}.\cite{bor_static_09}
We illustrate this method for $M_b=1$ first and generalize the results afterwards.

The eigenstates $|M_\nu\rangle$ can be classified by a set of $M_b+1$  
Bethe roots $\{\omega_{0,\nu},\ldots,\omega_{M_b,\nu}\}$ 
of the exact solution.  Their positions in the complex plane are determined 
by coupled non-linear 
equations.\cite{bor_inhom07,gaubook} The eigenvalues are given by 
\be
\L_{M_\nu}&=&-\frac12 \sum_{k=0}^{M_b}\omega_{k,\nu}+\frac{N_b x_1}{4}\,\;\;.\label{ev}
\ee
To each eigenstate belongs a distinct root pattern that is related to the flipped spins
relative to the {all up} state.\cite{bor_static_09}
In particular,
a Bethe root in the origin corresponds approximately to the application of a 
global lowering operator $S^-_{\rm tot}$, a 
root $A_{\ell+1}<\omega<A_{\ell}$ {induces essentially a superposition of states 
with spin flips} on sites $\ell$, $\ell+1$ \cite{footnote} and a 
root $\omega=\Or(N_b)$ mainly causes a flip of the central spin,  
 respectively.
Therefore, for $M_b=1$
the state {with the central spin and a single}
bath spin at site $\ell$
{in the down state}
is most strongly overlapping with the six eigenstates 
that are characterized by two Bethe roots as follows:
$|0,0\rangle$, 
$|0,\omega_{1,\ell-1}\rangle$, 
$|0,\omega_{1,\ell}\rangle$, 
$|0,\wt \omega_0\rangle$, 
$|\omega_{1,\ell}',\omega_{0,\ell}\rangle$, 
$|\omega_{1,\ell-1}',\omega_{0,\ell-1}\rangle$, 
where the  roots can be approximately determined from the distribution of $A_j$
in an expansion of $d:=(N_b\,x_1)^{-1}$ and $y_1:=d \sum_{j=1}^{N_b} A_j^{2}$ as
\cite{bor_static_09}
\be
\wt \omega_0&=& 1/d+y_1+\Or(d). \nonumber \\
\omega_{0,\ell}& =& 1/d+y_1-2A_\ell+ \Or (d) \label{roots}\\
\omega_{1,\ell}& \approx&\omega_{1,\ell}'= A_\ell+\Or(d) \nonumber
\ee
The sum of the squared overlaps $|\langle M_\nu|L\rangle|^2$ from only 
those states 
yields $0.71$ for $\ell=1$ increasing to $0.96$ for $\ell=N_b$ \cite{footnote}
in a system with $N_b=15$ and $x_1=B=2$ in
Eq.~\refeq{coup}. Therefore, most of the weight in the
expansion of $|L\rangle$ into eigenstates is indeed found by only considering 
the six states listed above. For more homogeneous 
couplings (B=0.4) the corresponding overlaps are significantly larger.

Once these most important contributing states are known, the actual values of the overlaps 
$\langle M_\nu|L\rangle$ are secondary, but the differences in the corresponding eigenvalues 
$\L_{M_\nu}$ 
determine the spectral distribution in Eqs.~(\ref{sz}) and (\ref{aljb})
and therefore the decay time.
Taking the corresponding differences of eigenvalues using Eqs.~(\ref{sz}-\ref{roots}), 
we find that 
the high-frequency contributions occur at $\Omega_0=\frac{1}{2d}\l(1+d y_1\r)$ and in an 
interval $\l[\Omega_{1,\ell}-\Delta_\ell/2,\Omega_{1,\ell}+\Delta_\ell/2\r]$ 
around the $\ell$-dependent frequency 
$\Omega_{1,\ell}=\frac{1}{2d}\l(1\!+\!d y_1\!-\!2 d A_\ell\r)$, where 
$\Delta_\ell:=(A_{\ell-1}-A_{\ell+1})/2$ up to terms of order $\Or(d)$. 
The long time scale $\sim 1/\Delta_\ell$ resulting from the width of the  
{peak near $\Omega_{1,\ell}$} can accordingly be
{interpreted} as the decoherence time for this dominant oscillation.
In addition, there are low frequency contributions around $\Omega_{2,\ell}=A_\ell/2 + \Or(d)$,
due to the inhomogeneity in the couplings $A_j$, which disappear for a homogeneous model.
\cite{mbjshom07}

Before turning to other polarizations, $M_b>1$, we can now also consider more realistic
 initial states with bath configurations {other than the $S_i^z$ 
product eigenstates hitherto considered.}
In the following we will use 
a uniform distribution over all bath states with a given polarization as the initial state, 
given by the density matrix 
\be
\rho_{\rm t}=\l({C_{M_b}^{N_b}}\r)^{-1}|\!\Down\rangle\langle\Down\!|\, \mathds{1}_{M_{b}}\label{rho0in},
\ee
where $\mathds{1}_{M_{b}}$ is the projection operator onto the sector 
with $M_b$ flipped 
bath spins relative to the {fully polarized up} state.  
In this case the {frequency spectrum of $\langle S_0^z \rangle (t)$} 
consists of the superposition of the spectra obtained 
for all allowed individual product states. Thus 
for $M_b=1$, all frequencies $\Omega_{1,\ell}$, $\Omega_{2,\ell}$ 
contribute, with $\ell=1,\ldots,N_b$. Thus the {high-frequency}
spectrum now consists of one sharp peak 
at $\Omega_0$ and of a second peak at $\Omega_1:= \Omega_0-x_1$
with a width $\Delta:=A_1-A_{N_b}$. 
Accordingly, 
the corresponding decoherence time for this oscillation is given by
\be
\tau=2\pi/\Delta\label{dectime},
\ee
which is therefore directly linked to the inhomogeneity of couplings.  
{Another way of 
interpreting these results is to say that the central spin 
{precesses}
in the effective field from the coupled bath state
with a relaxation time $T_1 \sim \tau$.}
The {low-frequency} tail contains frequencies in the interval $\l[A_{N_b},A_{1}\r]$.

\begin{figure}
\includegraphics[width=.99\columnwidth]{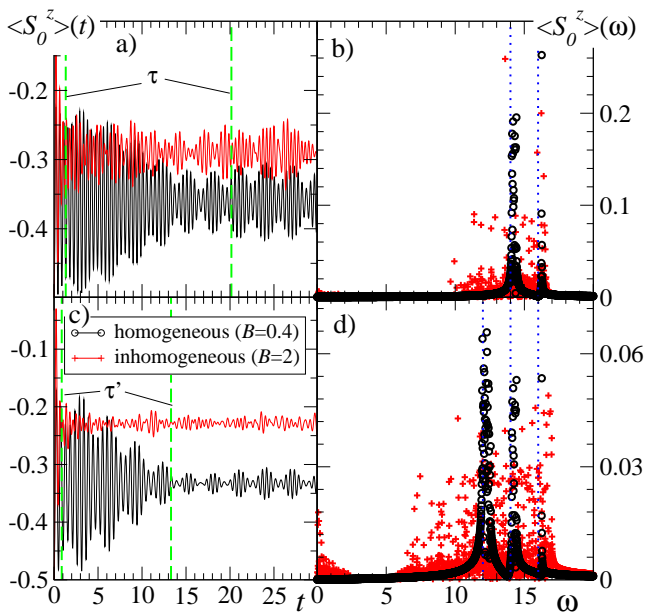}
\caption{ Time evolution $\langle S_0^z\rangle(t)$ (a,c) and the
corresponding Fourier transform (b,d) 
for $N_b=15$, $x_1=2$,  and an initial 
 uniform distribution of bath states with $M_b=1$ (in a,b) and $M_b=2$ (in c,d),
respectively.
The effect of two different homogeneity parameters $B=0.4$ (circles, black) and
$B=2$ (crosses, red) for the couplings in Eq.~\refeq{coup} is shown.
Dashed lines (green) show the estimates for the decoherence times. 
The dotted lines (blue) give the position of the peaks in the homogeneous limit {$A_j \equiv 2$}. }
\label{fig:dyn_n16}
\end{figure}

In order to illustrate the decoherence process 
we show the time evolution of $\langle S_0^z\rangle(t)$ and the corresponding 
Fourier transform $\langle S_0^z\rangle(\omega)$ 
for $N_b=15$ bath spins as obtained from complete diagonalization in Fig.~\ref{fig:dyn_n16}. 
We choose $x_1=2$ and two different values for $B$, 
corresponding to relatively homogeneous ($B=0.4$) and relatively inhomogeneous ($B=2$) couplings
in Eq.~\refeq{coup}. 
For nearly {homogeneous} couplings, the broadening of the 
{peak near $\Omega_1$} is demonstrated nicely. 
For $B=2$, the two peaks cannot be distinguished any longer due to the large broadening
of the peak around $\Omega_1$. 

We can generalize the above discussion to larger $M_b$ with the 
analogous initial density matrix in Eq.~\refeq{rho0in}. 
For $M_b=2$ three peaks are present 
 centered around $\Omega_{0}$ and $\Omega_{1}$ as given above, and 
around $\Omega_2:=\Omega_0-2x_1$.
Neglecting complex string solutions {of the Bethe Ansatz equations}  and 
interactions between excitations, one expects  
the peak near $\Omega_2$ to have twice 
{the }width $\Delta_2=2 \Delta$.
Again, in the homogeneous limit, the known \cite{bor_inhom07} frequencies $A(N_b+1-2k)$, $k=0,1,2$, are recovered.  From the beating of oscillations 
within the corresponding frequency ranges
we estimate the overall decoherence time to be 
$\tau^\prime=2 \tau/3$. 
The results for $M_b=2$ in Fig.~\ref{fig:dyn_n16} are consistent with 
the estimates for $\tau$ and $\tau^\prime$.
At the same time, one notices that additional spectral weight develops 
at small frequencies for increasing $M_b$ for inhomogeneous couplings. 
This can be traced back to the combination of states with Bethe roots 
$(0,\omega_{1,\ell}')$ and $(0,0)$ in the sum in Eq.~\refeq{sz}. 
Therefore, the smallest of these resulting frequencies is given by the most {\it weakly} 
coupled spins, $\Omega\sim A_{N_b}/2$, around which 
indeed most of the spectral weight in the low-frequency region of Fig.~\ref{fig:dyn_n16}-d
{develops}.  

It is possible to increase $M_b$ further and continue this analysis with more peaks
for {other 
 initial bath configurations} as long as $M_b \ll N_b$.
The positions and widths of the peaks directly reflect  the initially flipped bath spins 
and the choice of couplings.
For larger $M_b$, however, 
interactions between the elementary excitations may distort the simple analogy between roots
and flipped spins above. 
Therefore, we analyze the situation in the following
for small bath
magnetization $M_b\sim N_b/2$ 
using complete diagonalization on a smaller system in order to see which features of the 
analytic considerations can survive in that case.

\begin{figure}
\includegraphics[width=.99\columnwidth]{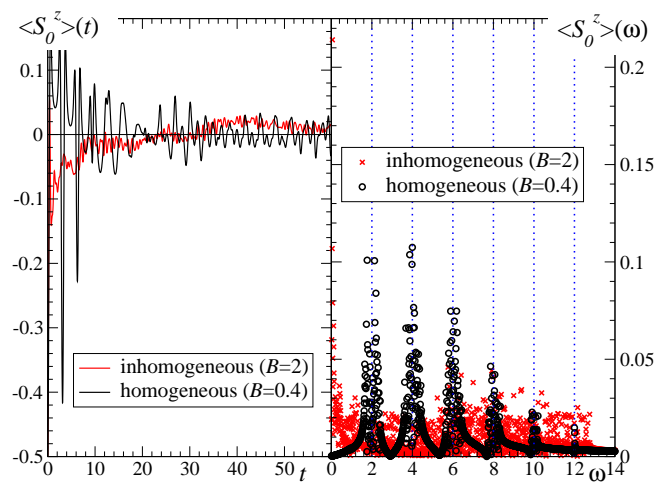}
\caption{ The time evolution $\langle S_0^z\rangle(t)$ (left) and its Fourier 
transform $\langle S_0^z\rangle(\omega)$ for $S^z_{\rm tot}=0$ from complete diagonalization,
for $N_b=11$,  $x_1=2$, and two different choices of the
homogeneity parameter $B$ in Eq.~\refeq{coup}. 
The dotted lines (blue) give the position of the peaks in the homogeneous limit $A=2$. }
\label{fig:dyn_n12}
\end{figure}

In Fig.~\ref{fig:dyn_n12}, we show results for $N_b=11$ 
in the subsector 
$S^z_{\rm tot}=0$, where we took the average over all allowed product states 
as in Eq.~\refeq{rho0in} with the central spin pointing down.
For a weak inhomogeneity, one can still clearly distinguish the discrete peaks in the Fourier transform, so that the analytical predictions are still useful for $M_b=5$ in this case. 
The decoherence times of the lower frequency oscillations are again generally shorter, as
can be seen by the widths of the corresponding peaks.

This structure is lost for couplings with a significant degree of inhomogeneity, 
{i.e. when the
difference in couplings becomes larger than the peak separation. }
However, as can be seen both in the time evolution and its Fourier transform, 
a new time scale at 
small energies occurs, which  leads to {low-frequency} oscillations. 
In the Fourier transform, this shows up as a {relatively strong peak at low frequency}. It is reasonable to expect that this stems from the same low-frequency 
mechanism as discussed above, namely 
the overlap between the ground state - with all roots in the origin except 
for $\omega_0$ - and the lowest excited state - 
where one root is shifted out of the origin into the interval 
between $A_{N_b}$ and $A_{N_b-1}$. This yields dominant  frequencies
corresponding to the most weakly coupled spins,  $\Omega\sim A_{N_b}/2$, 
leading to a characteristic long-time oscillation $4\pi/A_{N_b}$.
{The physical interpretation is that for generic disordered couplings 
fluctuations occur on all time scales up to the highest frequencies 
$\Omega_0 \sim \sum_j A_j$ leading to a correspondingly large $1/T_1$, 
but for longer times 
there remains a relatively coherent low-frequency 
oscillation,
 which emphasizes 
the importance of the weakly coupled nuclear bath spins in the time evolution 
of the central spin.\cite{che07}  This long time 
behavior can be explained by realizing that the
most weakly coupled spins simply play the role of a relatively stable backgound field, 
but never actually become strongly entangled with the rest of the system}.

In summary, we have analyzed the decoherence in the commonly used central spin model
using exact methods.  For relatively homogeneous couplings
the locations and widths of dominant oscillations
 in the Fourier spectrum 
$\langle S^z_0\rangle(\omega)$ 
can be predicted analytically
even for 
smaller polarizations (large $M_b$).
The positions and widths of the peaks directly reflect the initial bath state and the choice
of couplings.
High frequency oscillations have the longest decoherence times. 
For larger inhomogeneity in the couplings 
the decoherence times become shorter and eventually the 
simple peak structure is lost.
However, in this case the appearance
of a low frequency feature  
can again be inferred  from the root structure of the exact solution.
{Accordingly, it is possible to identify the physical behavior of the dynamics
in the different cases, namely fast oscillations with decoherence due to the 
difference in coupling strengths for nearly homogeneous couplings on the one
hand and relatively stable long time oscillations due to the most weakly coupled
background for inhomogeneous couplings on the other hand.}
The results also provide a direct link from clear signatures in the dynamics to individual 
excitations and characteristics of the model. 
This opens new possibilities  and insights for the analysis of the dynamical behavior
that can be obtained from independent methods, 
e.g.~from more advanced numerical or even experimental studies.

We are grateful to F.H.L.~Essler, F.~G\"ohmann, 
A.~Kl\"umper, I.~Schneider and A.~Struck for
useful discussions. M.B.~thanks the Rudolf-Peierls-Centre for Theoretical
Physics, University of Oxford, for kind hospitality. 
Financial support by the European network INSTANS and
the DFG via the research initiative SFB-TR49 is gratefully acknowledged. 

\providecommand{\bysame}{\leavevmode\hbox to3em{\hrulefill}\thinspace}
\providecommand{\MR}{\relax\ifhmode\unskip\space\fi MR }
\providecommand{\MRhref}[2]{%
  \href{http://www.ams.org/mathscinet-getitem?mr=#1}{#2}
}
\providecommand{\href}[2]{#2}


\begin{thebibliography}{10}
\bibitem{joh05}
A.~C. Johnson, J.~R. Petta, J.~M. Taylor, A.~Yacoby, M.~D. Lukin, C.~M. Markus,
  M.~P. Hanson, and A.~C. Gossard, Nature \textbf{435}, 925 (2005).

\bibitem{kha00}
A.V.~Khaetskii and Y.V.~Nazarov, Phys. Rev. B \textbf{61}, 12639 (2000);
{\it ibid.} \textbf{64}, 125316 (2001).


\bibitem{sch03}
J.~Schliemann, A.V.~Khaetskii, and D.~Loss, J. Phys.: Cond. Mat. \textbf{15}, R1809 (2003).  

\bibitem{jel04}
F.~Jelezko, T.~Gaebel, I.~Popa, A.~Gruber, and J.~Wrachtrup, Phys. Rev. Lett.
  \textbf{92}, 076401 (2004).

\bibitem{ber08}
J.~Berezovsky, M.~H. Mikkelsen, N.~B. Stoltz, L.~A. Coldren, and D.~D.
  Aschwalom, Science \textbf{320}, 349 (2008).

\bibitem{pre08}
D.~Press, T.~D. Ladd, B.~Thang, and Y.~Yamamoto, Nature \textbf{456}, 218 (2008).
  

\bibitem{han08}
R.~Hanson, V.~V. Dobrovitski, A.~E. Feiguin, O.~Gywat, and D.~D. Aschwalom,
  Science \textbf{320}, 352 (2008).

\bibitem{grei09}
A.~Greilich, S.~E. Economou, S.~Spatzek, D.~R. Yakovlev, D.~Reuter, A.~D.
  Wieck, T.~L. Reinecke, and M.~Bayer, Nat. Phys. \textbf{5}, 262. (2009)

\bibitem{rev} N.V.~Prokofev and P.C.E.~Stamp, Rep. Prog. Phys. {\bf 63}, 669 (2000).

\bibitem{kha02}
A.V.~Khaetskii, D.~Loss, and L.~Glazman, Phys. Rev. Lett. \textbf{88}, 186802 (2002);
Phys. Rev. B \textbf{67}, 195329 (2003).  


\bibitem{sch02}
J.~Schliemann, A.V.~Khaetskii, and D.~Loss, Phys. Rev. B \textbf{66} 245303 (2002).
  

\bibitem{dob03}
V.~V. Dobrovitski and H.~A. De~Raedt, Phys. Rev. E \textbf{67}, 056702 (2003).


\bibitem{coi04}
W.~A. Coish and D.~Loss, Phys. Rev. B \textbf{70}, 195340 (2004).

\bibitem{erl04}
S.~I. Erlingsson and Y.~V. Nazarov, Phys. Rev. B \textbf{70} (2004), 205327.

\bibitem{has06}
K.~A. Al-Hassanieh, V.~V. Dobrovitski, E.~Dagotto, and B.~N. Harmon, Phys. Rev.
  Lett. \textbf{97} 037204, (2006). 

\bibitem{coi06}
W.~A. Coish, E.~A. Yuzbashyan, B.~L. Altshuler, and D.~Loss, J. Appl. Phys.
  \textbf{101}, 081715 (2007).

\bibitem{den06}
C.~Deng and X.~Hu, Phys. Rev. B \textbf{73}, 241303(R) (2006).

\bibitem{che07}
G.~Chen, D.~L. Bergman, and L.~Balents, Phys. Rev. B \textbf{76},  045312 (2007).
 

\bibitem{gaubook}
M.~Gaudin, \emph{La fonction d'onde de {B}ethe}, Masson, 1983.

\bibitem{bor_inhom07}
M.~Bortz and J.~Stolze, Phys. Rev. B \textbf{76}, 014304 (2007).

\bibitem{bor_static_09}
M.~Bortz, S.~Eggert, and J.~Stolze,  Phys. Rev. B {\bf 81}, 035315 (2010). 


\bibitem{mbjshom07}
M.~Bortz and J.~Stolze, J. Stat. Mech. , P06018 (2007).

\bibitem{koz07}
G.~G. Kozlov, JETP \textbf{105}, 803 (2007).

\bibitem{caux09}
A.~Faribault, P.~Calabrese, and J.-S. Caux, J. Stat. Mech., P03018 (2009).

\bibitem{footnote}
{The correspondence of roots to local spin flips  
is particularly good for the more loosely bound spins $\ell\gg 1$.}




\end{thebibliography}
\end{document}